\documentclass[a4paper]{article}
\usepackage{tipa}
\usepackage{INTERSPEECH2019}
\usepackage{hyperref}
\title{A CNN-based tool for automatic tongue contour tracking in ultrasound images}
\name{Jian Zhu$^1$, Will Styler$^2$, Ian Calloway$^1$}
\address{
  $^1$Department of Linguistics, University of Michigan,  United States\\
  $^2$Department of Linguistics, University of California San Diego, United States}
\email{\{lingjzhu,iccallow\}@umich.edu,wstyler@ucsd.edu}

\begin{document}

\maketitle
\begin{abstract}
For speech research, ultrasound tongue imaging provides a non-invasive means for visualizing tongue position and movement during articulation. Extracting tongue contours from ultrasound images is a basic step in analyzing ultrasound data but this task often requires non-trivial manual annotation. This study presents an open source tool for fully automatic tracking of tongue contours in ultrasound frames using neural network based methods. We have implemented and systematically compared two convolutional neural networks, U-Net and Dense U-Net, under different conditions. Though both models can perform automatic contour tracking with comparable accuracy, Dense U-Net architecture seems more generalizable across test datasets while U-Net has faster extraction speed. Our comparison also shows that the choice of loss function and data augmentation have a greater effect on tracking performance in this task. This public available segmentation tool shows considerable promise for the automated tongue contour annotation of ultrasound images in speech research.
\end{abstract}

\section{Introduction}
Ultrasound tongue imaging provides a non-invasive means for assessing tongue position and movement during speech production.  However, the presence of speckle noise and irrelevant high contrast edges often degrades the usability of ultrasound images by obscuring the tongue surface \cite{stone_guide_2005}. Consequently, extracting tongue contours from ultrasound images remains a non-trivial task. 

In linguistic and clinical phonetics, extracting tongue contours is the usually first step in analyzing ultrasound images, but this process is time-consuming. For phoneticians and speech scientists, tongue contours offer a direct visualization and measurement of certain articulatory processes. In the past decade, various methods for semi-automatic or automatic tongue contour extraction have been proposed to facilitate the analysis of ultrasound data, notably the Active Contour (Snake) based methods \cite{li_automatic_2005,xu_robust_2016,laporte2018multi}, graph based methods \cite{tang_graph-based_2010}, and neural network based methods \cite{jaumard-hakoun_tongue_2016,berry_dynamics_2011,fabre_tongue_2015,wen2018automatic,zhu2018automatic,mozaffari2019bownet}. Both Snake based and graph based methods are mostly semi-automatic, which still require manual initialization, but methods like automatic initialization \cite{xu_robust_2016} or particle filtering \cite{laporte2018multi} can gear the algorithm towards more automatic segmentation. Neural network based methods are promising for fully automatic segmentation. Prior works utilized deep neural networks \cite{berry2010automatic,fabre2017automatic} and Boltzmann machines \cite{jaumard-hakoun_tongue_2016}; recently fully convolutional neural networks such as variants of the U-Net \cite{ronneberger2015u} have been adapted to segment tongue contours \cite{wen2018automatic,zhu2018automatic,mozaffari2019bownet,mozaffari2019transfer}. 

Studies comparing some of the publicly available methods show that semi-automatic or automatic tracing can approximate human annotations under some conditions \cite{xu_comparative_2016,csapo_error_2015}, but these tools require either extensive human intervention, relevant technical knowledge, or proprietary software, which considerably limit their usage. Few studies explore the generalizability of contour tracking methods across speakers and different ultrasound machines \cite{mozaffari2019transfer}.

In this paper, we extend previous works on U-Net based models \cite{wen2018automatic,zhu2018automatic} by implementing a new tool for automatic tongue contour extraction using both U-Net and Dense U-Net. We systematically tested the performance of these models with different test datasets. The results show that, while both U-Net and Dense U-Net can achieve high accuracy in automatic tracking, loss function and data augmentation have a larger impact on actual tracking performance. In this task, the deeper Dense U-Net might not necessarily outperformed the shallower U-Net if not properly trained. Most importantly, given that a fully automatic tool for contour tracking is not public available, we are filling this gap by releasing a new open source tool to facilitate the otherwise time-consuming process of contour tracking in speech production research.

\section{Method}
In our approach, we first train a convolutional neural network to segment the brightest edge corresponding to the tongue tissue-air interface from a noisy ultrasound image, and then derive a tongue surface curve through post-processing of the segmented image. The source code, pre-trained models and some of the test data are available at \url{https://github.com/lingjzhu/mtracker.github.io}.

\subsection{U-Net}
\begin{figure}
    \centering
    \includegraphics[width=\linewidth]{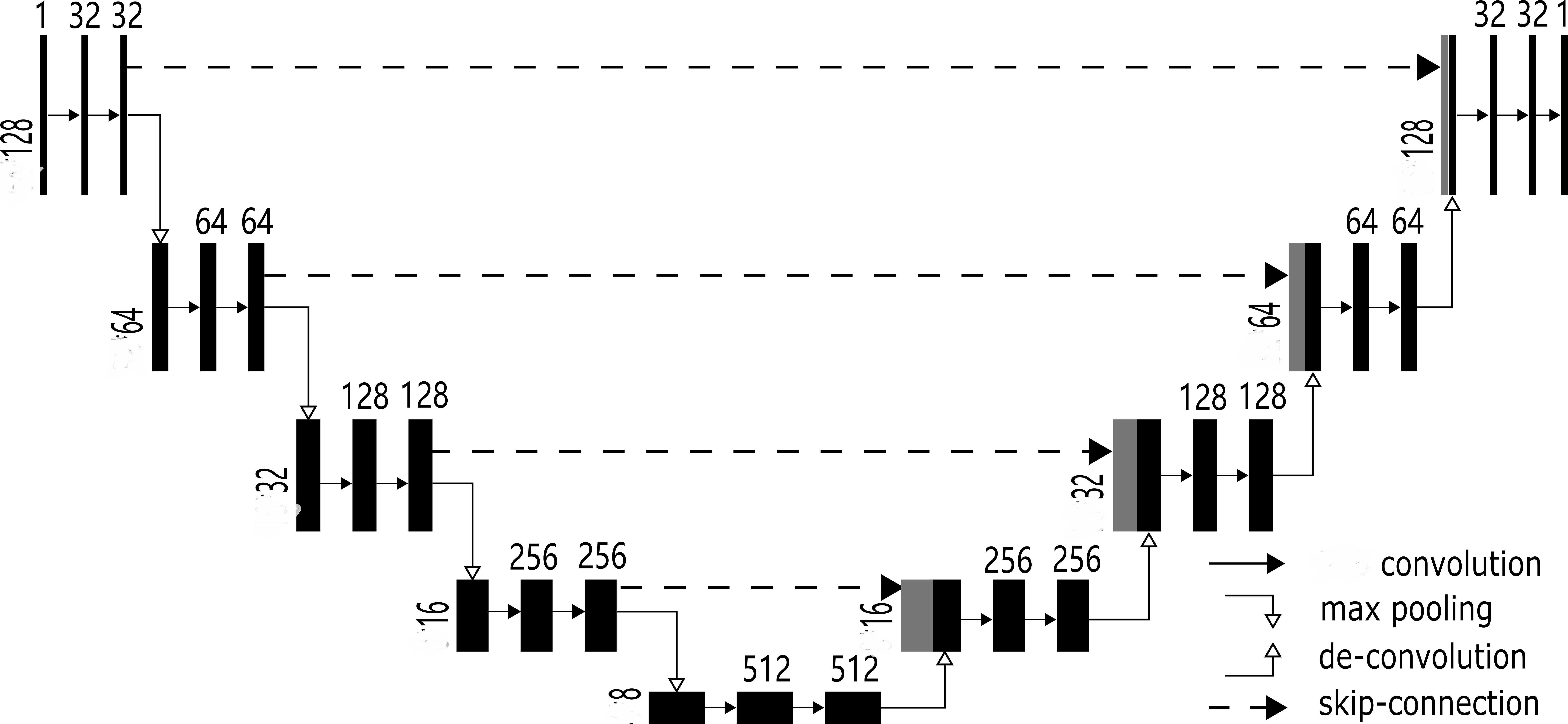}
    \caption{The U-Net architecture. Each rectangle represents the output feature maps and arrows represent different operations. The vertially displayed number to the left of rectangles indicates the image size at that block (e.g., the vertally displayed 128 stands for an image size of 128 $\times$ 128). The number at the top presents the number of feature maps, or image channels.}
    \label{fig:unet}
\end{figure}

For the baseline, we have adopted the U-Net architecture, a variant of the Fully Convolutional Neural Network (FCNN) widely used in medical image segmentation \cite{ronneberger2015u}. The typical U-Net architecture consists of a downsampling path with repeated convolution blocks and max-pooling layers, and an upsampling path with deconvolution layers and convolutional blocks (see Fig. \ref{fig:unet}). U-Net also introduced the skip-connection, or concatenating feature maps in the downsampling path and feature maps in the upsampling path to enable the reuse of low-level features in higher layers. We used the following settings for the U-Net model. Each convolutional block has the following components: 3$\times$3 conv + rectified linear unit (ReLU) + 3$\times$3 conv + ReLU + 2$\times$2 max pool.  Each de-convolutional block in the upsampling path has 2$\times$2 up-conv + 3$\times$3 conv + ReLU + 3$\times$3 conv + ReLU. All convolution operations use a stride of one and zero padding. The number of feature maps doubled after each convolutional block with a range of (32, 64, 128, 256, 512), and halved after each de-convolutional block with a range of (256, 128, 64, 32). The final layer was a 1$\times$1 conv layer with sigmoid activation. 

\subsection{DenseNet and Dense U-Net}
The Dense Convolutional Network (DenseNet) is a network architecture that has been shown to be effective in many computer vision tasks, outperforming some of the classic architectures such as ResNet \cite{huang2017densely}. Dense U-Net is an adapted network architecture that fuses both DenseNet and U-Net, thereby adapting DenseNet to segmentation task at the pixel level \cite{li2018h,guan2018fully}. It combines the DenseNet and the U-Net by introducing a symmetric upsampling path and long range skip-connections to enable the reuse of low-level features. 

In this study, we adopted the standard DenseNet-121 architecture \cite{huang2017densely} as the downsampling path by removing its top classification layer, leaving only the dense blocks and transition layers. Each dense block has repeated convolutional blocks consisting of batch normalization (BN) + ReLU + 1$\times$1 conv + BN + ReLU + 3$\times$3 conv with a growth rate of 32, or the number of feature maps of each convolution layer. There are 6, 12, 24 and 16 convolutional blocks in four dense blocks respectively. Within the dense block, the input feature maps feeds into a sequence of operations mentioned above, which produces the output feature maps.  Then the input and output feature maps are concatenated together to become the input for next sequence of operations.  For the transition layers, each consists of BN + ReLU + 1$\times$1 conv + 2$\times$2 average pool. 

In the upsampling path, de-convolutional layers are used to increase the image size and skip-connnecting the corresponding dense blocks with later layers allows us to reuse the feature maps, as in U-Net. Each de-convolutional block has a 2$\times$2 de-convolutional layer and a dense block. Each dense block in the upsampling path has a single convolutional block (BN + ReLU + 1$\times$1 conv + BN + ReLU + 3$\times$3 conv) with 16, 24, 12, 6 and 6 growth rates respectively. As de-convolutional layers can also perform feature extraction alongside upsampling, each dense block only has a single convolutional sub-block. Finally, the output layer is a 1$\times$1 conv layer with sigmoid activation, which is used to resize and scale the feature maps to a single-channel grayscale image. 

\begin{figure}[t]
  \centering
  \includegraphics[width=\linewidth]{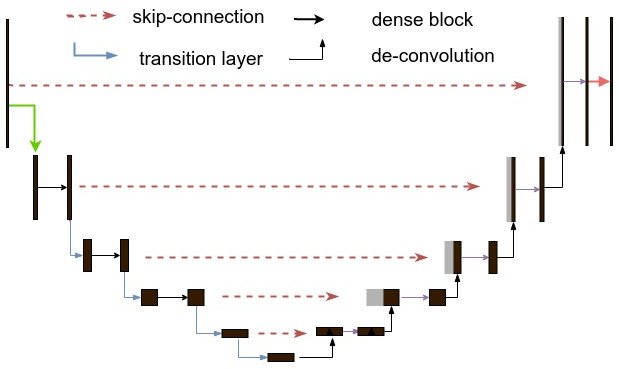}
  \caption{The structure of a Dense U-Net, which consists of multiple downsampling and upsampling dense blocks. This structure is highly similar to U-net. However, the original U-Net's max pooling layers are replaced by transition layers and the U-Net's convolution blocks are replaced by dense blocks.}
  \label{fig:dunet}
\end{figure}

\subsection{Loss functions}
One of the main challenges in this task was the extreme class imbalance between tongue-related pixels and the irrelevant background pixels. On average, the relevant pixels corresponding to a tongue-shape annotation (a `mask') only comprise 2\% of the total pixels. Different loss functions have been proposed to address this question. The Dice Similarity Coefficient (DSC) \cite{milletari2016v} only penalizes the mismatch between the predicted white pixels (representing the tongue region) and the white edge in the mask, while excluding all background pixels and noise during the optimization process. Thus, the learning task can be formulated as minimizing the following loss function:

\begin{equation}
\mathcal{L}\textsubscript{DSC} = -\frac{2\sum_{i=1}^{N}s\textsubscript{i}r\textsubscript{i}+\epsilon}{\sum_{i=1}^{N}s\textsubscript{i}+\sum_{i=1}^{N}r\textsubscript{i}+\epsilon} 
\end{equation}

where $s\textsubscript{i}$ is the softmax output between 0 and 1, and $r\textsubscript{i}=1$ when i is in the ground truth contour and 0 otherwise. $s\textsubscript{i}\in S$ represents the predicted tongue region given by the CNN and $r\textsubscript{i}\in R$ the ground truth. A smoothing factor of $\epsilon$, which was set to 1 here, was added to to make the loss function smooth and to avoid zero division. Compared with WSC, although the DSC can generate a slim tongue spline, but it also tends to force the model to generate probability values close to either 0 or 1, leading to overconfidence. The generated heatmaps are highly binarized, which is not a good reflection of the probablistic encoding of the original masks.

Another way to counterbalance the disparity between two classes is to use the weighted binary crossentropy loss \cite{xie2015holistically}. Assigning too large of a weight to the minority class (contour) may cause the model to overpredict the minority class, resulting in oversmoothing the predicting tongue shape. Given the standard crossentropy loss,

\begin{equation}\label{dice}
    \mathcal{L}\textsubscript{C} = - \sum_{i=1}^{N}y_i\log(p_i) + (1 - y_i)\log(1 - p_i)
\end{equation}

The class weighted crossentropy (Eq. \ref{eq:wc}) assigns different weights $w_p$ and $w_n$ to the two categories by setting the weights to be the inverse of the ratios of two categories respectively. 

\begin{equation}\label{eq:wc}
    \mathcal{L}\textsubscript{WC} = - \sum_{i=1}^{N}w_py_i\log(p_i) + w_n(1 - y_i)\log(1 - p_i)
\end{equation}

The compound loss (Eq. \ref{eq:com}) is the weighted sum of the Dice loss and the standard crossentropy loss, with the weight $\lambda$ being a hyperparameter that can be tuned. The standard crossentropy functions as a regularizer to control the overconfidence given by DSC, forcing the model to generate a more gradient probabilistic heatmap.

\begin{equation}\label{eq:com}
    \mathcal{L}\textsubscript{Compound}=\mathcal{L}\textsubscript{DSC}+\lambda*\mathcal{L}\textsubscript{C}
\end{equation}

By adjusting $\lambda$, we can tune the predicted heatmap.  We set $\lambda=5$ in the current task based on pilot experiments with validation data. In order to assess the effect of these loss functions, we systematically compared the performance of three loss functions, namely the Dice loss, the weighted crossentropy (WC) and the compound loss. 
%\iffalse

%\fi

\section{Data}
Midsagittal ultrasound data was collected as MPEG video at 60 frames per second, using a Zonare Z.One Ultrasound Unit, operating at 4MHz and 70Hz scan rate with a P4-1C transducer. Tongue shape curves were annotated with Mark Tiede's GetContours package for MATLAB \cite{getcontour} \footnote{\url{https://github.com/mktiede/GetContours}}, generating a 100 point spline for each curve from human-specified anchor points. Annotators were trained to mark the bottom edge of the white reflectance signal corresponding to the tongue surface. Our data consisted of 35160 human-annotated ultrasound frames from 11 American English speakers producing vowel and vowel-lateral syllable nuclei in C\textipa{2l}C and C\textipa{2}C pairs (e.g. `bulk' and `buck'), collected for another project. 

The data were split into training, validation and test sets through random partitioning, each consists of 45\%, 5\% and 50\% of the total data. All models were trained only on the training dataset. In order to test the generalizability of our model to multiple machines and configurations, we used three datasets, listed below. Except the NS test data, the remaining test sets were manually annotated by the first author. All images were scaled to 128 $\times$ 128 pixels.

\begin{itemize}
    \item The NS test data consisted of 3926 frames from two additional American English speakers (one male and one female) reading `The North Wind and the Sun', collected using the same equipment and settings as the training data, but annotated in its entirety by each of three trained annotators.
    \item The Ultrax test data consisted of 793 ultrasound images collected from a male typical developing child and a female child with speech disorder \cite{eshky2018ultrasuite}. 
    \item The UltraSpeech test data were primarily 241 ultrasound frames from two French sentences, each read by a different male French speaker \cite{fabre2017automatic}.
\end{itemize}

\begin{figure}
  \centering
  \includegraphics[width=0.45\linewidth]{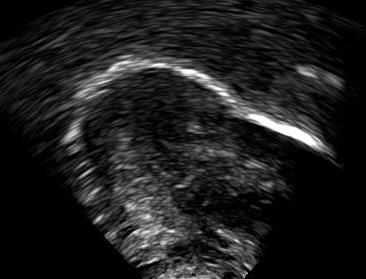}
  \includegraphics[width=0.45\linewidth]{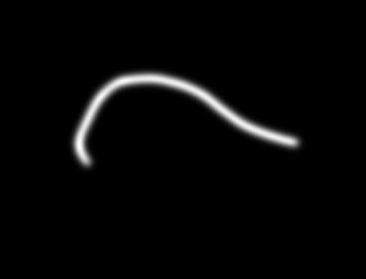}
  \caption{A sample ultrasound frame and its corresponding mask.}
  \label{fig:sample}
\end{figure}

\subsection{Masks}
Each human annotation is represented by 100 pairs of Cartesian x-y coordinates. Each tongue-shape annotation was generated as a probability heatmap of the same size as the original ultrasound image  (`mask'). Given a sequence of x-y coordinates $[(x_1,y_1),(x_2,y_2),...,(x_n,y_n)]$, the Gaussian kernel in Eq. \ref{eq:g} was used to map the human-created 100-point tongue contour data into the mask. The $I(x,y)$ indicates the pixel intensity at point (x,y), representing the probability of each pixel being part of the tongue contour. Thus, pixels closer to the actual tongue surface coordinates are assigned higher probabilities, while all other pixels are gradually diminishing to 0 as they are further away from the contour. The key is to treat each point as the center of a Gaussian distribution and then create a distribution over it on the mask. Then the distribution for each point is added up and then normalized between [0,1].  

\begin{equation}\label{eq:g}
I(x,y) 
\propto  \sum_{i=1}^{N}exp\Big(-\frac{(x-x_i)^2+(y-y_i)^2}{2\sigma^2}\Big)
\end{equation}
%\sum_{i=1}^{N}\frac{1}{2\pi\sigma^2}exp\Big(-\frac{(x-x_i)^2+(y-y_i)^2}{2\sigma^2}\Big) \\
The $I(x,y)$ indicates the pixel intensity at point (x,y), representing the probability of each pixel being part of the tongue contour. Thus, pixels closer to the actual tongue surface coordinates are assigned higher probabilities, while all other pixels are gradually diminishing to 0 as they are further away from the contour. In actual implementation, the default $\sigma$ in this study is set to 4, and values below 0.4 were thresholded to only retain pixels with high probabilities.

\section{Experiments}
The training data were divided into multiple mini-batches, each with a size of 32 images. We used the Adam optimizer \cite{kingma2014adam} with a learning rate of 0.0001, and the model was trained for 30 epochs. The training process took approximately 2 hours using an NVIDIA Tesla K40 GPU in the University of Michigan's FLUX computing cluster. The model that achieved the lowest validation loss was retained as the final model.

\subsection{Post-processing}
For each new image fed into the model, the output is a probability heatmap having the same size as the input image, with the intensity of each pixel again corresponding to the probability that the pixel is part of the tongue. A 50\% threshold is then applied to the image to filter out unlikely predictions. Then a skeletonization algorithm \cite{zhang1984fast} is used to reduce the white edge to a single pixel wide representation. It is then interpolated and smoothed using 'UnivariateSpline' in the SciPy Package with the default settings. The resulting output is a 100-point Cartesian coordinate representation of the predicted tongue shape. 

\section{Evaluation}
The metric for evaluation of error from human annotation is the Mean Sum of Distance (MSD), which permits the comparison of two curves without requiring point-wise alignment \cite{li_automatic_2005}. The MSD between two sequences U and V can be computed as the average distance between a given point and its nearest point in another sequence:

\begin{equation}
D(U,V)=\frac{1}{2n}\Bigg(\sum_{i=1}^{n}\min_{j}|v\textsubscript{i}-u\textsubscript{j}|+\sum_{i=1}^{n}\min_{j}|u\textsubscript{i}-v\textsubscript{j}|\Bigg)
\end{equation}

where n is the number of points in each sequence, and u\textsubscript{i} and v\textsubscript{j} are pairs of x-y coordinates from two sequences U and V under comparison. 

\subsection{Same-speaker evaluation}
Table \ref{tab:same} below displays the average MSD between our model and the human annotators on all 17580 test ultrasound images\footnote{We attempted to run comparisons with prior splining algorithms, but we were unable to find an appropriate set of hyperparamters for TongueTrack \cite{tang_tongue_2012} for our dataset, and were unable to run AutoTrace \cite{fasel_deep_2010} because of deprecated dependencies.}. With the exception of the weighted crossentropy loss, other models performed almost equally well, achieving an MSD as small as 0.85mm (about 3.5px). The low tracking error is likely due to the fact that these test images were from the same group of speakers in the training data.

\begin{table}[th]
  \caption{Mean and (Standard Deviation) of Mean Sum of Distance (in Pixels, 1 pixel $\approx$ 0.25mm) for the 17580 frame test dataset.}
  \label{tab:same}
  \centering
  \begin{tabular}{lr}
    \toprule
Model                 & \multicolumn{1}{l}{MSD}   \\ \midrule
UNet\textsubscript{\textbf{-WC}}             & 4.26 (2.13)                   \\
UNet\textsubscript{\textbf{-Dice}}           & 3.42 (2.05)                   \\
UNet\textsubscript{\textbf{-Compound}}       & 3.53 (2.05)                    \\
D UNet\textsubscript{\textbf{-WC}}       & 4.35 (2.06)                \\
D UNet\textsubscript{\textbf{-Dice}}     & \textbf{3.25 (1.96)}                 \\
D UNet\textsubscript{\textbf{-Compound}} & 3.79 (2.20)                  \\ 
    \bottomrule
  \end{tabular}
\end{table}

\subsection{Cross-speaker evaluation}
Table \ref{tab:nw} shows the model performance on the NS test dataset relative to each human annotator. The average human-to-human difference (measured using MSD) was around 0.7mm (an estimated 2.79px), whereas the average-human-to-CNN difference was around twice that (1.25mm, 5px). This remains good performance, though, particularly given that the NS test set contains two speakers and more diverse tongue shapes corresponding with fluid speech. The Dense U-Net with compound loss shows slightly better performance relative to the other models but, given the large variance resulting from outliers, it cannot be conclude that model architecture matters in this task. 

The results also demonstrate that weighted crossentropy is not a suitable loss function for the current task as models trained with weighted crossentropy lagged considerably behind other models. In the weighted crossentropy loss, the "tongue contour" class is given a much higher weight, so the model tends to predict as much "tongue contour" class as possible to minimize the loss, resulting in thick contours in prediction. In contrast, the Dice loss might be more effective in dealing with class imbalance. The downside of the Dice loss is that it biases the model to predict the "tongue contour" class with a probability that is uniformly 1, leading to overconfidence. In the compound loss (Eq.\ref{eq:com}), the standard crossentropy term is similar to a regularization term. As the Dice loss only assesses the intersection between the ``the tongue contour'' class in the prediction and the ground truth, the crossentropy term puts some weights on those ``background'' pixels in the masks, resulting in a more gradient prediction. It turned out that this loss function can reduce more outlier predictions than the Dice loss function. 

\begin{table}[th]
  \caption{Mean and (Standard Deviation) of Mean Sum of Distance (in Pixels, 1 pixel $\approx$ 0.25mm) for the NS test set, as compared to three human annotators A, B and C.}
  \label{tab:nw}
  \centering
  \begin{tabular}{lrrr}
    \toprule
     & \multicolumn{1}{l}{A} & \multicolumn{1}{l}{B} & \multicolumn{1}{l}{C}  \\ \midrule
     A & 0 (0) & 2.33 (1.57) & 2.83 (1.85) \\
     B & 2.33 (1.57) & 0 (0) & 3.21 (2.21)\\ 
     C & 2.83 (1.85) & 3.21 (2.21) & 0 (0) \\ \midrule
UNet\textsubscript{\textbf{-WC}}             & 6.65 (2.92)           & 6.44 (2.74)           & 7.25 (3.24)            \\
UNet\textsubscript{\textbf{-Dice}}           & 5.70 (2.68)           & 5.33 (2.37)           & 6.09 (2.87)           \\
UNet\textsubscript{\textbf{-Compound}}       & 5.31 (2.60)           & 4.93 (2.25)           & 5.64 (2.76)            \\
D UNet\textsubscript{\textbf{-WC}}       & 7.74 (3.27)           & 7.48 (3.03)           & 8.25 (3.45)            \\
D UNet\textsubscript{\textbf{-Dice}}     & 5.15 (2.54)           & 4.77 (2.19)           & 5.65 (2.68)            \\
D UNet\textsubscript{\textbf{-Compound}} & \textbf{5.01 (2.52)}           & \textbf{4.58 (2.12)}           & \textbf{5.33 (2.63)}           \\ 
    \bottomrule
  \end{tabular}
\end{table}

\subsection{Data augmentation and training data size}
\begin{figure}[t]
  \centering
  \includegraphics[width=\linewidth]{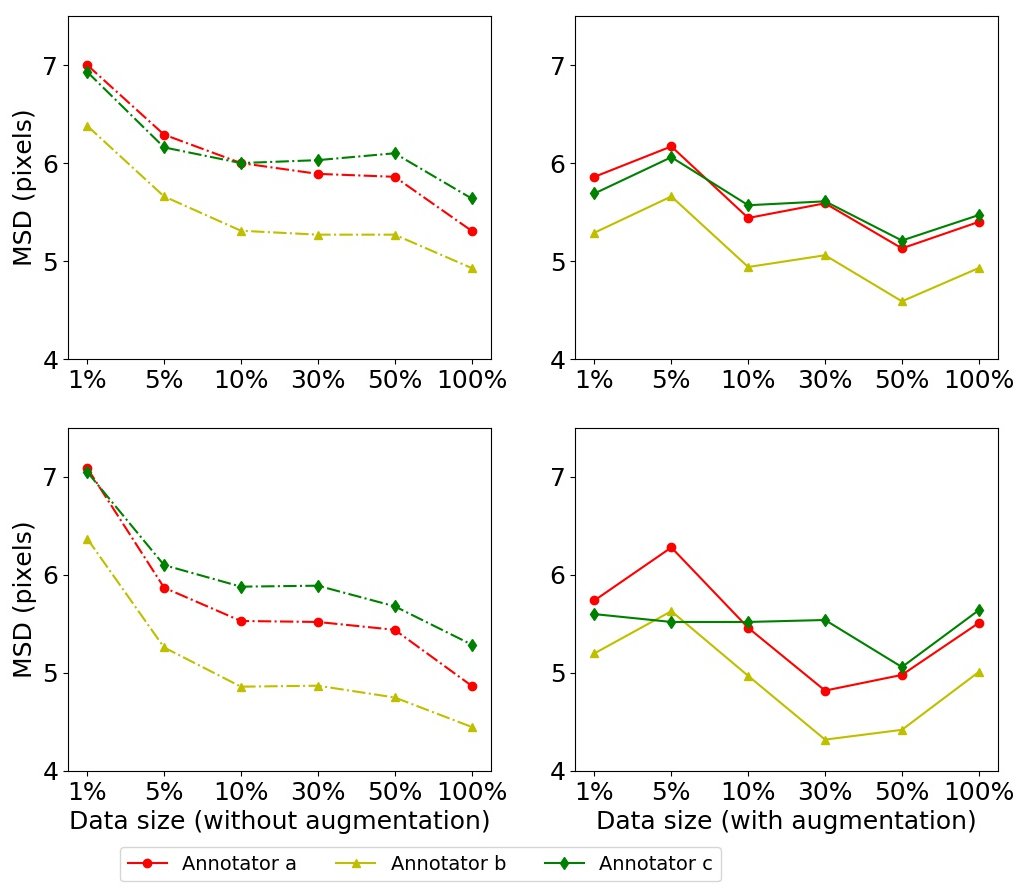}
  \caption{The influence of data size on tracking performance in NS dataset. The upper panels show results of U-Net models with varying training data size, while the lower panels display results from Dense U-Net models. Solid and dashed lines represents results under with-augmentation or without-augmentation conditions, respectively.}
  \label{fig:size_aug}
\end{figure}

We retrained models by varying the training data size incrementally from 1\%, 5\%, 10\%, 30\%, 50\% to 100\%. Data augmentation was applied to generate more diverse training data, including horizontal flipping, rotation within the range of $-15^{\circ}$ and $+15^{\circ}$, zooming, and horizontal and vertical shifting.  Fig.\ref{fig:size_aug} demonstrates that, despite some minor fluctuations, MSD tends to decrease with more training data. Models with data augmentation outperformed models without augmentation by a small margin when the original training data size was small, but the improvement brought by augmentation disappeared when the model was trained with more training data. It is noticeable that Dense U-Net consistently showed slight improvements over U-Net with the original training data. However, after training with augmented data, both U-Net and D-Net tend to have very similar performance, even approximating the best performance with as small as 30\% of the entire training data (around 5000 frames). With data augmentation, the difference in performance resulting from difference in architecture was greatly diminished. Even with only 1\% of training data (about 160 frames) the model can achieve reasonable accuracy with data augmentation. This highlights the importance of data augmentation in the current task.

\subsection{Image size}
Image size also affects the model performance, as shown in Table \ref{tab:size}. Though both models produced comparable MSD with image size of 32 $\times$ 32 or 224 $\times$ 224,  Dense U-Net gives the lowest MSD when the image size is 64 $\times$ 64. These results show that images with more details might not necessarily result in improved performance.

\begin{table}[th]
  \caption{\label{tab:table1} Mean and (Standard Deviation) of Mean Sum of Distance (in Pixels, 1 pixel $\approx$ 0.25mm) for the NS test dataset averaged over three annotators.}
  \label{tab:size}
  \centering
  \begin{tabular}{lrrr}
    \toprule
Model                 & \multicolumn{1}{l}{32 $\times$ 32} & \multicolumn{1}{l}{64 $\times$ 64} & \multicolumn{1}{l}{224 $\times$ 224}  \\ \midrule
UNet\textsubscript{\textbf{-Compound}}  & 5.81 (2.86)           & 5.94 (3.05)           & 5.66 (2.66)            \\
D UNet\textsubscript{\textbf{-Compound}} & \textbf{5.60 (2.93)}           & \textbf{5.03 (2.52)}           & \textbf{5.53 (2.78)}           \\ 
    \bottomrule
  \end{tabular}
\end{table}

\subsection{Testing images from different machines}
To test the generalizability of our model, we examined the model performance with the additional Ultrax and UltraSpeech datasets, the results of which are displayed in Table \ref{tab:different}. All models can well predict the tongue shapes in the Ultrax test set with similar performance. The UltraSpeech test set posed a bigger challenge to these models because its noise distributions are quite different from those in the training data, but the Dense U-Net with compound loss proved to be stable across dataset, as it produces low MSD in both test sets. However, these results should also be interpreted with caution as the accuracy is sensitive to the selection of ROI. We used the same ROI for all models, but the results may vary if a different ROI is selected. This again highlights the difficulty of cross-domain prediction. A potential solution to cross-domain prediction can be transfer learning \cite{mozaffari2019transfer}.

\begin{table}[th]
  \caption{\label{tab:table2} Mean and (Standard Deviation) of Mean Sum of Distance (in Pixels) for different test datasets.}
  \label{tab:different}
  \centering
  \begin{tabular}{lrr}
    \toprule
Model                 & \multicolumn{1}{l}{Ultrax}  & \multicolumn{1}{l}{UltraSpeech}   \\ \midrule
UNet\textsubscript{\textbf{-Dice}}           & \textbf{5.52 (1.65)}           & 7.92 (4.74)                    \\
UNet\textsubscript{\textbf{-Compound}}       & 6.10 (3.36)           & 8.25 (4.83)                      \\
D UNet\textsubscript{\textbf{-Dice}}     & 6.35 (2.40)           & 6.68 (3.46)                     \\
D UNet\textsubscript{\textbf{-Compound}} & 5.71 (1.66)           & \textbf{5.72 (2.88)}                     \\ 
    \bottomrule
  \end{tabular}
\end{table}

\begin{figure}[t]
  \centering
  \includegraphics[width=0.3\linewidth]{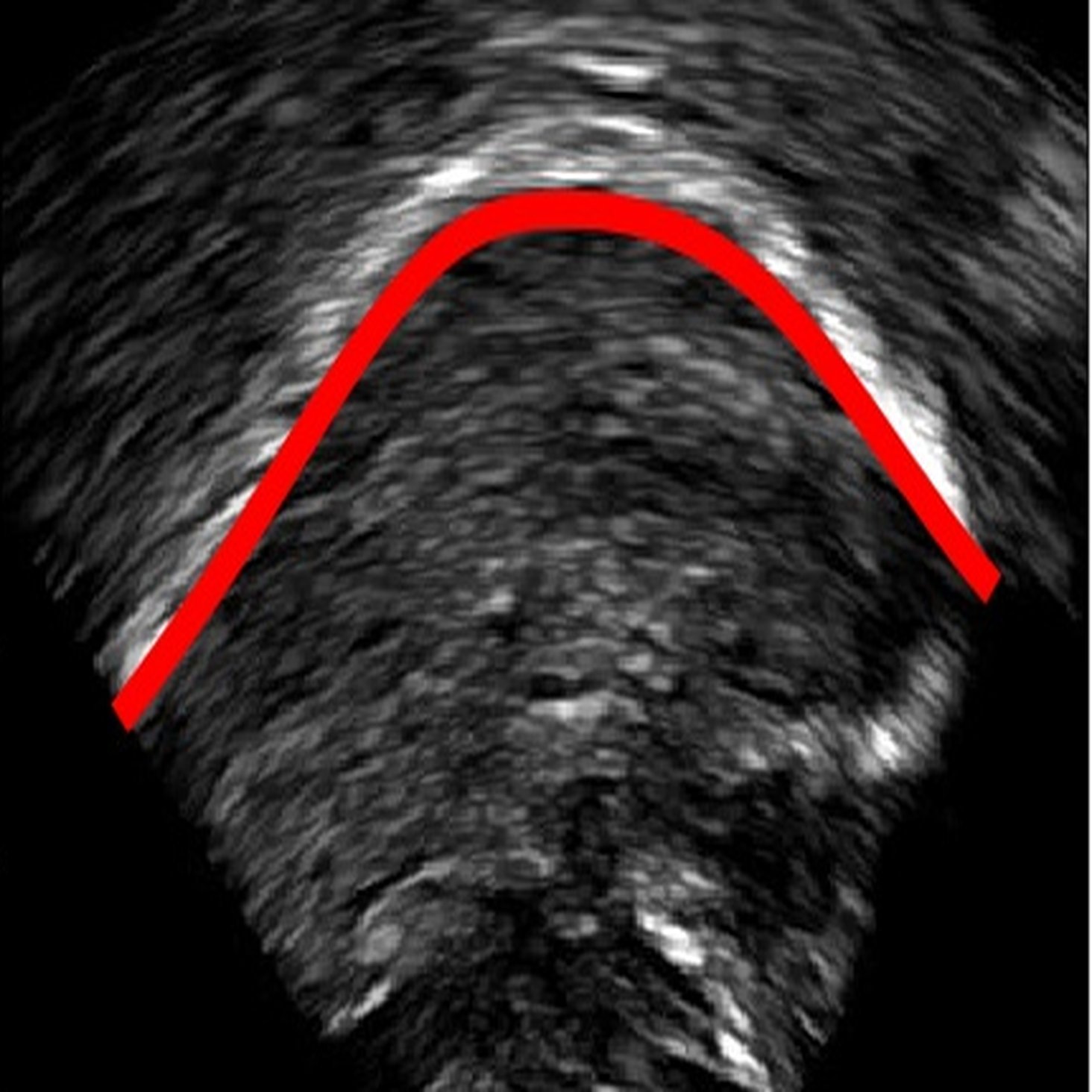}
  \includegraphics[width=0.3\linewidth]{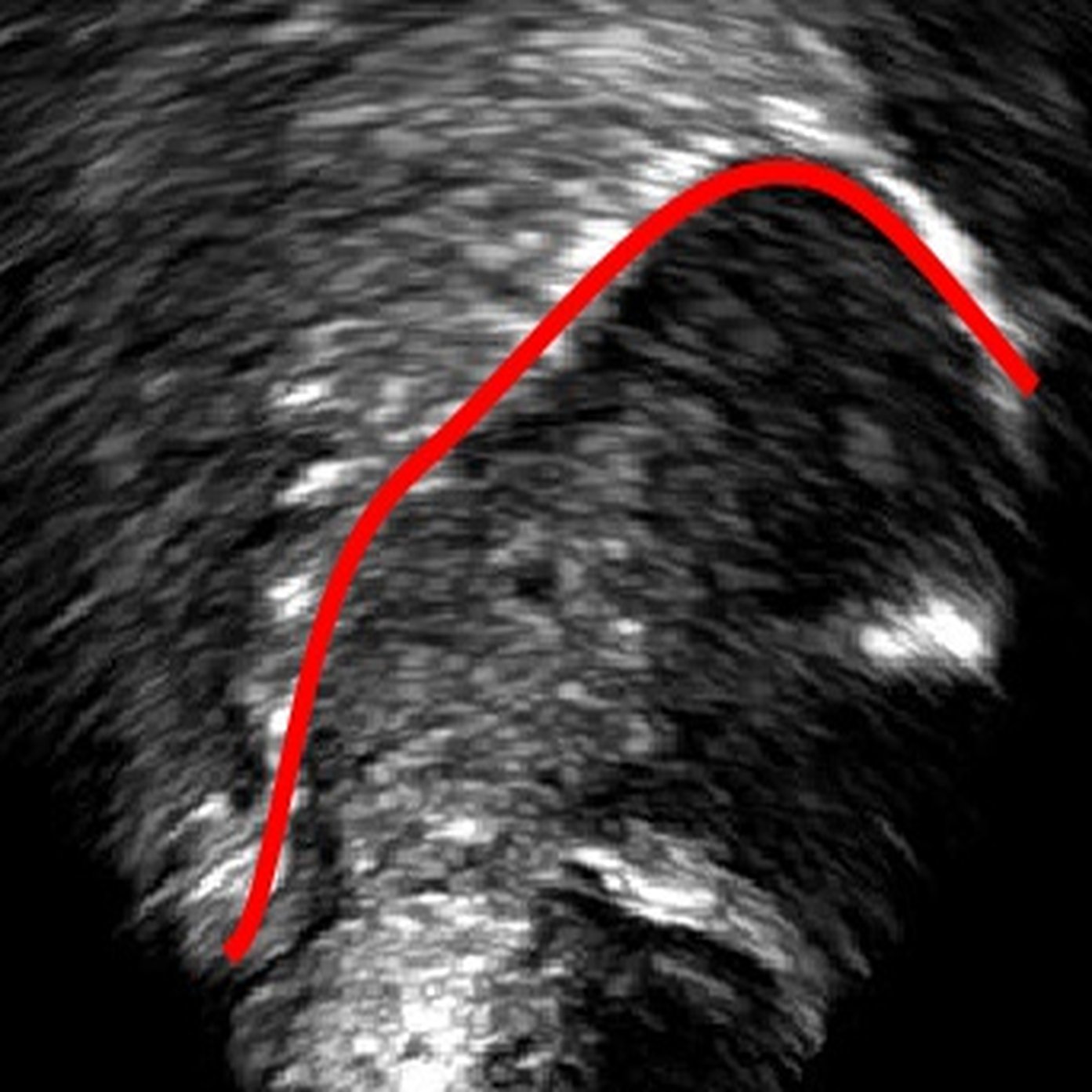}
  \includegraphics[width=0.3\linewidth]{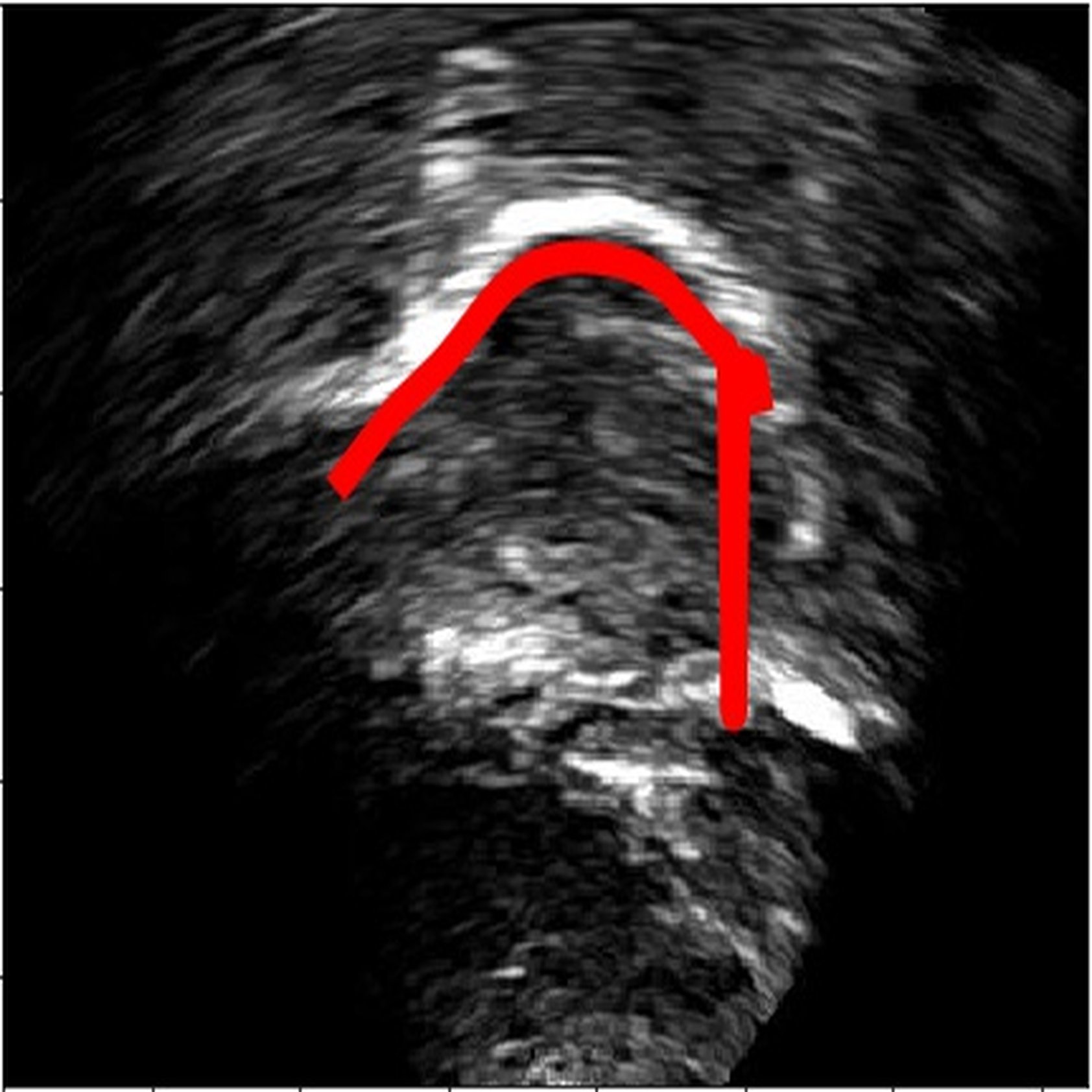}
  \includegraphics[width=0.3\linewidth]{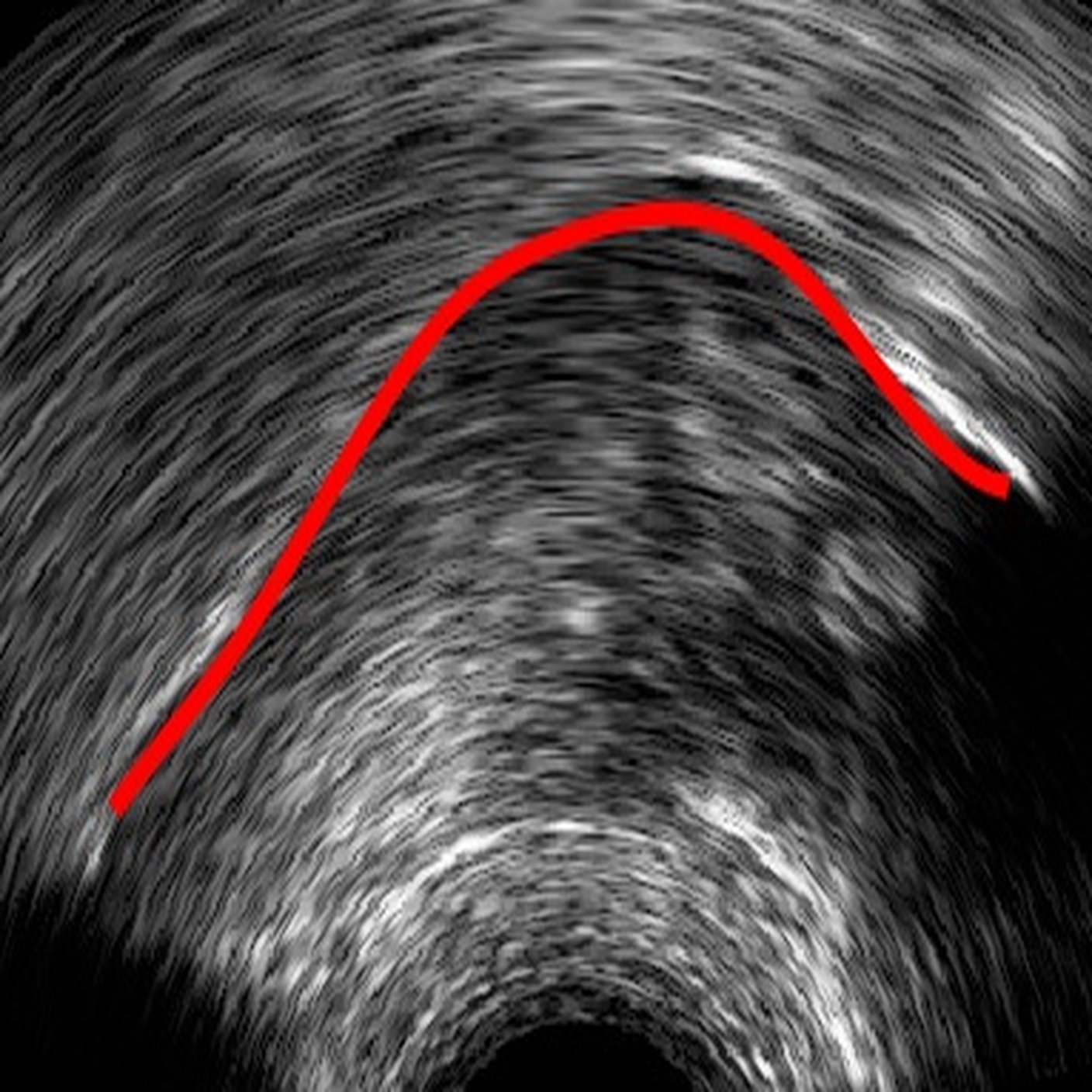}
  \includegraphics[width=0.3\linewidth]{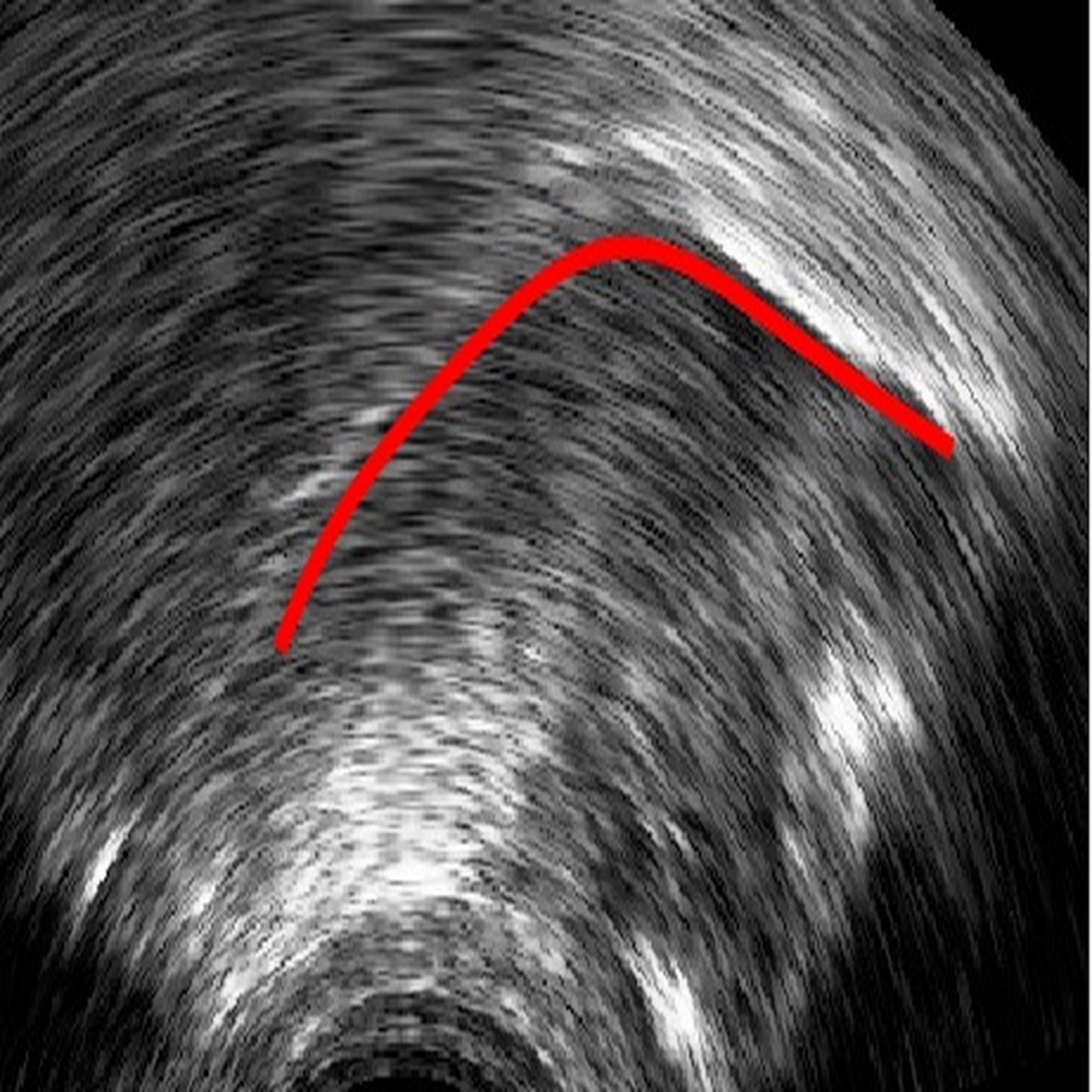}
  \includegraphics[width=0.3\linewidth]{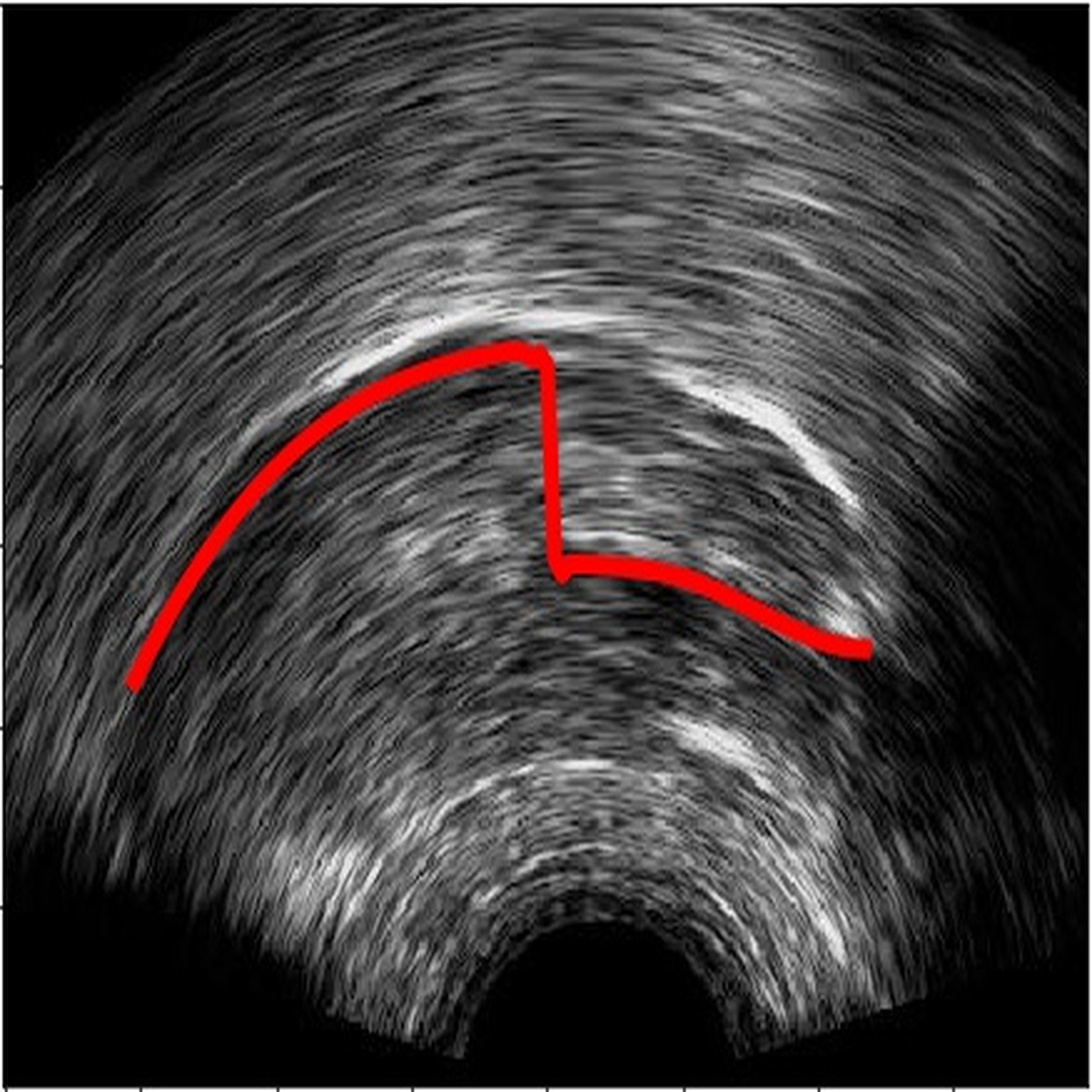}
  \caption{Sample predictions given by D U-Net\textsubscript{-Compound}. Upper panels are from the NS test set; lower panels are from the UltraSpeech test set. The speckle noises in ultrasound images sometimes can lead to failures in identifying parts of the tongue surface.}
  \label{fig:test}
\end{figure}

\section{Error analysis}
As the CNN is trained to identify the white edges directly corresponding to the tongue surface, additional or missing white edges due to bad image quality or speaker physiology can lead to failures in identifying parts of the tongue surface. In the absence of prior knowledge of plausible tongue shapes, the model will sometimes generate tracking errors when the white edge becomes blurry or interrupted. Similarly, bright edges in the image background are likely to be recognized as part of the tongue; tongue contours generated from image frames with these edges will likely suffer from implausible curvatures as interpolation in post-processing attempts to connect these regions. There some potential solutions to these problems, including incorporating temporal constraints on tongue contour variations across frames \cite{xu_robust_2016}, or adding a smooth constraints that penalizes discontinuity of tongue contours, or introducing a strong prior probability of possible tongue locations. In data processing, these issues can also be mitigated by tuning the parameters in post-processing to match the needs of the specific dataset, and remaining errors can also be addressed through manual correction (as even then, the workload is considerably reduced relative to manually labeling all frames). 

\section{Conclusions}
In this study, we present a new open source tool for fully automated tongue contour extraction based on U-Net and Dense U-Net models. The implemented models are tested extensively on multiple test datasets. Though both models can perform automatic contour tracking with comparable accuracy, Dense U-Net architecture seems more generalizable across datasets but U-Net has faster extraction speed. Our evaluation results show that the choice of loss function and data augmentation have a larger effect on model performance than simply stacking more layers. Crucially, unlike many prior solutions, our tool requires minimal human intervention to obtain point-by-point splines. The average speed for U-Net is $\sim$63 frames per second, and $\sim$29 frames per second for Dense U-Net on a consumer-grade laptop with Intel i-5 8600K processors and Nvdia 1070Ti GPU. The automatic contour extraction performed by our tool can potentially facilitate the time-consuming manual annotations in phonetic and clinical research. 

\section{Acknowledgements}
We are grateful to Patrice Speeter Beddor, Andries Coetzee, Thomas Hueber and the UltraSuite research group for making available their ultrasound data. The data from Beddor and Coetzee were collected for a different project supported by NSF grant BCS-1348150.

\bibliographystyle{IEEEtran}
\bibliography{template}

% \begin{thebibliography}{9}
% \bibitem[1]{Davis80-COP}
%   S.\ B.\ Davis and P.\ Mermelstein,
%   ``Comparison of parametric representation for monosyllabic word recognition in continuously spoken sentences,''
%   \textit{IEEE Transactions on Acoustics, Speech and Signal Processing}, vol.~28, no.~4, pp.~357--366, 1980.
% \bibitem[2]{Rabiner89-ATO}
%   L.\ R.\ Rabiner,
%   ``A tutorial on hidden Markov models and selected applications in speech recognition,''
%   \textit{Proceedings of the IEEE}, vol.~77, no.~2, pp.~257-286, 1989.
% \bibitem[3]{Hastie09-TEO}
%   T.\ Hastie, R.\ Tibshirani, and J.\ Friedman,
%   \textit{The Elements of Statistical Learning -- Data Mining, Inference, and Prediction}.
%   New York: Springer, 2009.
% \bibitem[4]{YourName17-XXX}
%   F.\ Lastname1, F.\ Lastname2, and F.\ Lastname3,
%   ``Title of your INTERSPEECH 2019 publication,''
%   in \textit{Interspeech 2019 -- 20\textsuperscript{th} Annual Conference of the International Speech Communication Association, September 15-19, Graz, Austria, Proceedings, Proceedings}, 2019, pp.~100--104.
% \end{thebibliography}

\end{document}